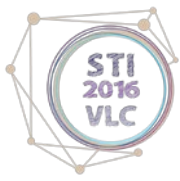

# Tracing scientific mobility of Early Career Researchers in Spain and The Netherlands through their publications[1]

Nicolas Robinson-Garcia[*], Carolina Cañibano[*], Richard Woolley[*] and Rodrigo Costas[**]

[*]elrobin@ingenio.upv.es; cacaisan@upvnet.upv.es; ricwoo@ingenio.upv.es
INGENIO (CSIC-UPV), Universitat Politècnica de València, Camino de Vera s/n, 46022 Valencia (Spain)

[**] rcostas@cwts.leidenuniv.nl
Centre for Science and Technology Studies (CWTS), Leiden University, Leiden (The Netherlands)

## INTRODUCTION

International scientific mobility is acknowledged to be a key mechanism for the diffusion of knowledge, particularly tacit or 'sticky' knowledge that cannot be transferred without geographical proximity and personal contact (Frenken 2010), for the incorporation of young researchers into elite transnational scientific networks (Laudel 2005), and for accessing additional resources or infrastructures that are essential to the research process but located in other places. The benefits of mobility as a means for knowledge diffusion are used as justification for the support policies and programmes that encourage the mobility of European Researchers (CEC 2004, 2008). The demand for researcher mobility indicators has been building in the EU since the launching of the Mobility Strategy for the European Research Area (CEC 2001).

The inadequacy and lack of appropriate data to assess the phenomenon of researcher mobility has been repeatedly pointed out by scholars and policy makers (CEC 2004; Fontes 2007; Didou-Aupetit 2009; Ackers 2009). Traditional comparable data sources such as migration or labour statistics and censuses allow us to picture flows of highly skilled human resources (HRST) (Cañibano & Woolley 2015) but not the movements of researchers which is more complex and often circulatory (Jöns 2007). However, important progress in the understanding of the phenomenon has been made building on the direct collection of information from researchers via surveys (IDEA Consult 2013), interviews (Ackers 2009) and CV analysis (Cañibano et al, 2011).

From a bibliometric perspective, Laudel (2003) suggests using the address field of publications to identify mobility patterns between researchers. While this approach seems feasible for studies based on small samples, it has been problematic at a large scale as researchers must be identified univocally. Author disambiguation is a historically problematic issue in bibliometric research when applied to individual scholars (Smalheiser & Torvik, 2009). One of the major limitations has been the lack of reliable methods for the proper identification (i.e. overcoming the usual problems of homonyms and synonyms found in personal names) of the full scientific output of the authors (Costas & Bordons, 2009; Reijnhoudt et al., 2014).

---

[1] Nicolas Robinson-Garcia is currently supported by a Juan de la Cierva-Formación Fellowship from the Spanish Ministry of Economy and Competitiveness.

In recent years, two different methods have been proposed for dealing with this problem. One is the Author ID developed by Scopus; the other is the algorithm developed by Caron and van Eck (2014). The development of such methodologies, together with the incorporation of the linkages between authors and affiliations in the records covered in two of the most important bibliographic databases (Web of Science and Scopus) has allowed the exploration of individual address changes in publications at a large scale. In this regard, Moed and colleagues have worked on the potential of bibliometric data as a proxy to measure mobility using the Scopus Author ID (Moed, Aisati, & Plume, 2013; Moed & Halevi, 2014),

This paper builds on their work and presents an exploratory analysis of different typologies of researchers according to their traceable mobility using scientific publications covered in the Web of Science (WoS). We compare two populations of researchers, of the same 'scientific age', based in Spain and The Netherlands. We address the following research questions:

1. Are there substantial differences in terms of research publication and citation impact between mobile researchers and non-mobile researchers within each country?

2. When does mobility take place in the research trajectory of a researcher? For those who return to their original country, how many years does it take?

3. Are there differences between the mobility profiles, publication productivity and citations of the Spain-based and Netherlands-based researchers?

The next section details the data and methods of analysis used and the following section specifies the results. The final section includes discussion of the results and of contextual factors associated with the Spanish and Dutch research systems that may contribute to explaining the initial outcomes we observe.

## MATERIAL AND METHODS

To establish reasonable comparisons between researchers based in The Netherlands and Spain, we must first identify similar groups of researchers in each country. We only consider 'trusted' direct linkages between author and affiliation as reported in scientific publications and recorded by WoS. In other words, we don't expect false positives in our linkages (i.e. wrong linkages between authors and affiliations) although we may have false negatives (i.e. we may miss linkages between the authors and the affiliations that have not been recorded in the publications). We define researchers' scientific age based on the year when they published their first reported paper with a target affiliation. As we aim at analysing career trajectories, we focus on researchers who published their first paper between 2003 and 2005. This way we ensure that they are all of a similar 'generation' (Costas, Nane & Larivière, 2015). We then define their home country as the one to which they were affiliated in their first year of publication. We work only with researchers who published their first paper whilst affiliated to an organization in either Spain or The Netherlands. Based on this set of researchers, we build their full publication record up to 2014. That way we can analyse their full publication career. We establish three different study groups:

*Mobile versus non-mobile researchers.* We consider a researcher as mobile when at any point in time during their career they have declared more than one country in the address field of one or more of their publications.

*Returned versus not returned researchers.* We consider a researcher has returned to their home country when they have reported within their later publications that they are again affiliated within the same country in which they published their first paper, following a period in which they were affiliated in a different country. This does not mean that it is the actual 'home' country of the researcher, rather it is their country of scientific origin as derived from their publication profile. In cases of double affiliation in The Netherlands and Spain, we have considered them as belonging to both countries. That is, a small sample of our data is duplicated for each country.

*Single versus multiple affiliation.* One of the limitations we found in our study is the role of researchers who report an affiliation to more than one institution or country. For that reason, we decided to analyse these special cases as a separate group.

These three study groups are not distinct from one another. In fact, the two latter are overlapped subgroups of the mobile researchers group. At this point we did not cross different variables as we wanted to conduct a first analysis of the dichotomous distinction between the groups and make comparisons between the two countries.

## RESULTS

Our data set includes a total of 10,412 researchers. Table 1 shows their distribution among the three classes of study groups. More than 20% of researchers from The Netherlands reported some degree of international mobility. Spain shows a lower international mobility rate (13.4%). However, while only 39.7% of mobile researchers from The Netherlands (n=344) had 'returned', the share increases up to 60.8% of mobile researchers from Spain (n=501). With regard to multiple affiliations, the share of researchers who reported being affiliated to institutions in more than one country at the same time is higher in the case of The Netherlands (9.81% of the total share) than for Spain (7.35%).

Table 1. Descriptive indicators of the study groups and home country

|  | SPAIN | | THE NETHERLANDS | |
|---|---|---|---|---|
|  | **N** | **%** | **N** | **%** |
| **Non-mobile researchers** | 5327 | 86.60 | 3394 | 79.65 |
| **Mobile researchers** | 824 | 13.40 | 867 | 20.35 |
| Returned researchers | 501 | 8.15 | 344 | 8.07 |
| Multiple affiliation | 452 | 7.35 | 418 | 9.81 |
| **Total** | **6151** | **100.00** | **4261** | **100.00** |

Figure 1 compares research productivity in terms of publications for the three classes of study groups and for The Netherlands and Spain. The boxplot on the left compares non-mobile with mobile researchers. Mobile researchers tend to have higher levels of production than non-mobile researchers. Differences are very small in the case of The Netherlands but slightly more pronounced for Spain. The middle boxplot compares returned vs. not-returned researchers. Here we find larger differences between the two groups, both for Spain and The Netherlands. Returned researchers show higher levels of publications production than those mobile researchers who have not returned home. Finally, at the right is the comparison between researchers with a single affiliation and those with multiple affiliations. An almost identical distribution is apparent in the case of The Netherlands and a slight difference in distribution is evident for Spain, although the average number of publications per researcher in Spain is very similar regardless of the number of affiliations.

Figure 1. Boxplot on publications distribution by study group and home country

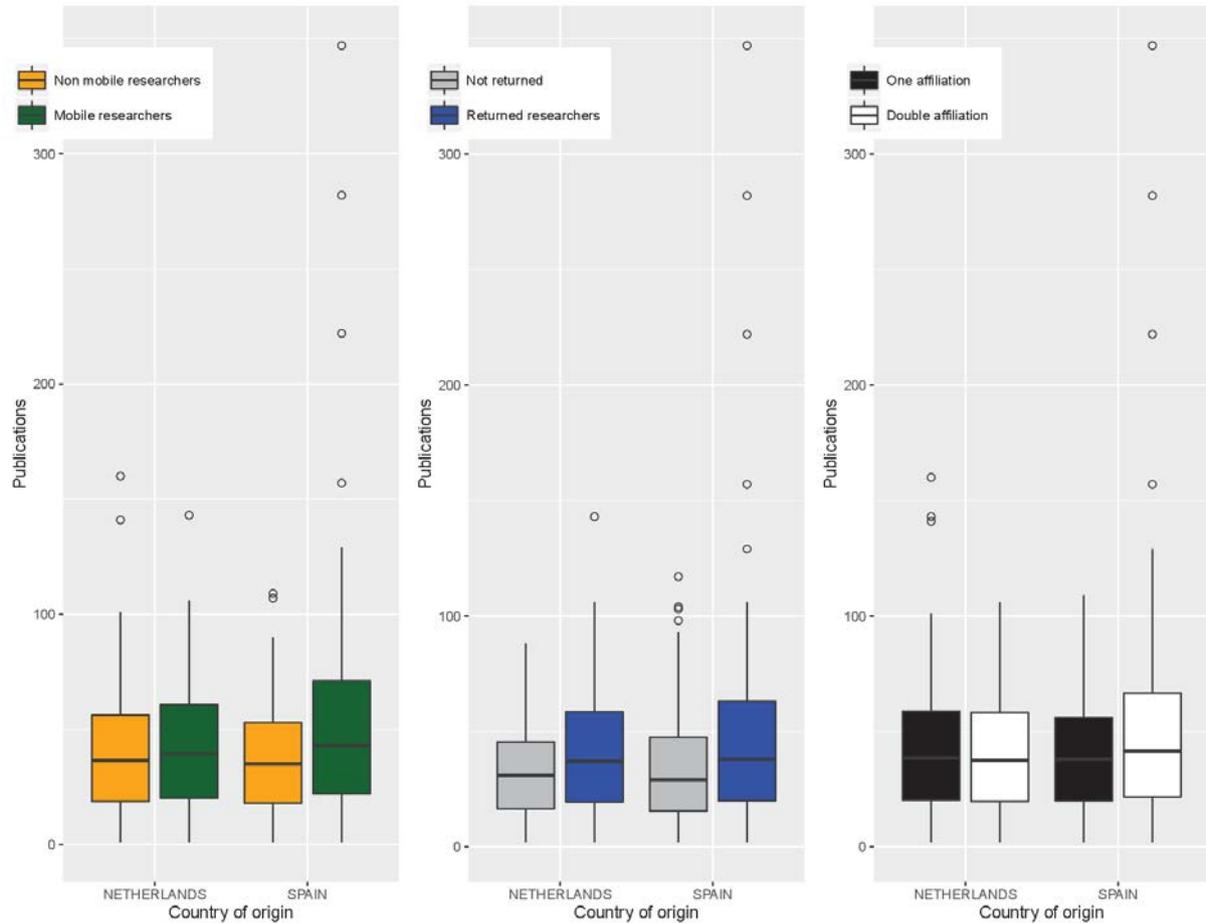

With regard to the relationship between the type of researcher and citation impact, we relate the share of highly cited papers in comparison with the total number of publications. Figure 2 compares each researcher type by country. In the case of the Netherlands, there seems to be no difference in the citation impact of researchers depending on their mobility and affiliation attributes. (However it should be noted that data is sparser for returned versus not-returned researchers and for researchers with multiple affiliations.)

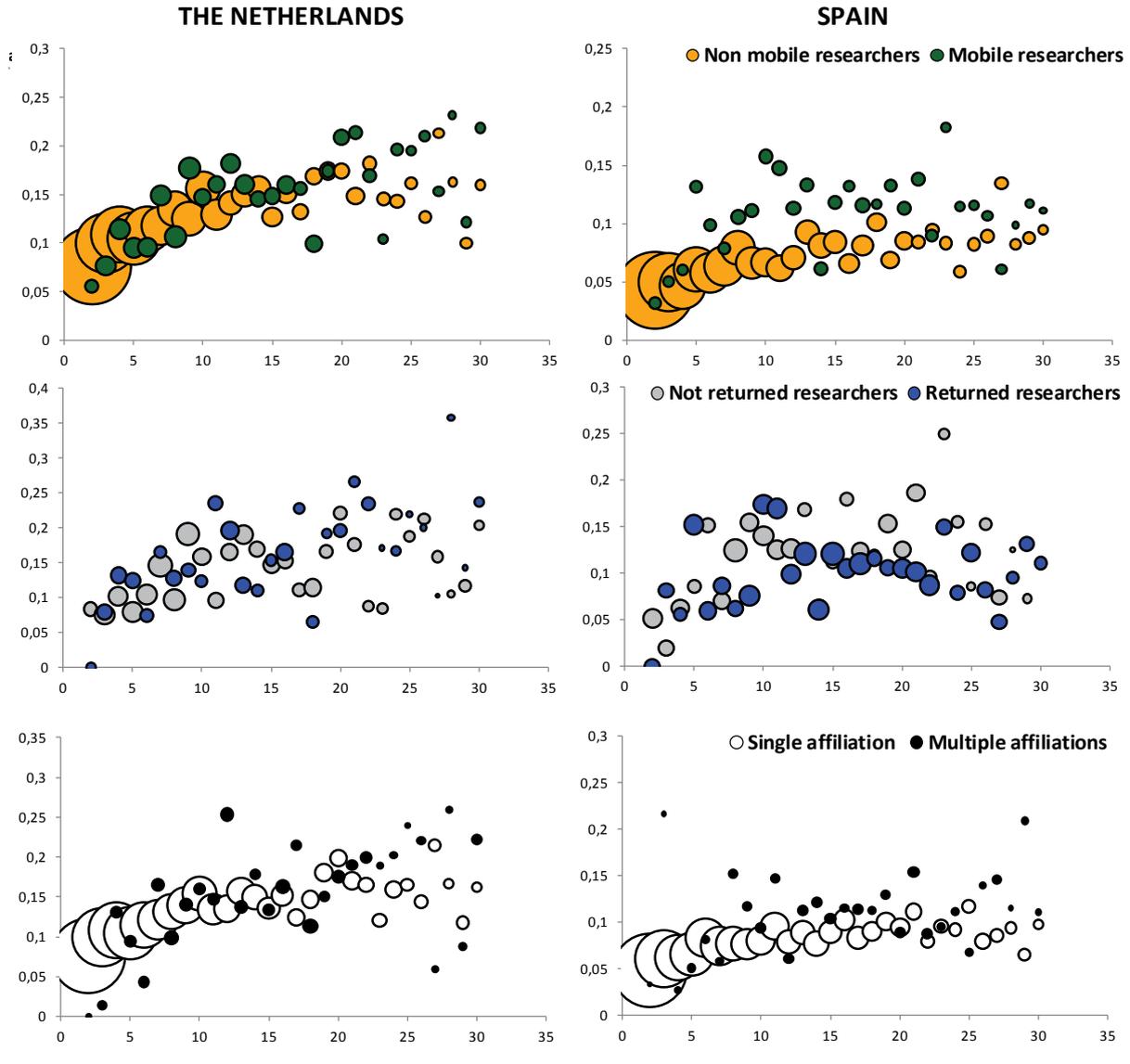

Figure 2. 10% Highly Cited Papers and publications by study group and home country. Size represents number of researchers. Y axis represents share of highly cited papers, X axis represents number of publications. Only included, researchers with >2 and <31 publications

In contrast, we do find differences in the Spanish case for the three classes of groups, showing higher impact for mobile, not-returned and multiple affiliation researchers. Again, we find that the relation between publication output and citation impact is not as correlated in the case of returned, not-returned and double affiliation researchers.

Table 2 analyses two different temporal aspects: number of years that it takes for mobile researchers to leave their home country; and number of years reported abroad for returned researchers. Here again we see differences between the Spain and The Netherlands. Researchers assigned to Spain take a longer period until they leave their home country (4.6 years on average compared to 4.2 years for researchers assigned to The Netherlands) and also stay abroad for a shorter period (around 3 years on average compared to around 4 years for The Netherlands).

Table 2. Years to emigration for mobile researchers and years to return for returned researchers

| NETHERLANDS | Years to emigration | Years abroad | SPAIN | Years to emigration | Years abroad |
|---|---|---|---|---|---|
| Average | 4.17 | 3.80 | Average | 4.61 | 2.98 |
| Std. Dev. | 2.51 | 3.09 | Std. Dev. | 2.51 | 2.77 |
| Population | 867 | 344 | Population | 824 | 501 |

# DISCUSSION AND FURTHER RESEARCH

We observe differences in the degree of mobility of Spain and Netherlands based researchers. The results are based on evidence of mobility that is long enough to lead to a change in affiliation, such as a job change or extended research visits. This does not necessarily reflect the overall degree of mobility of researchers based in Spain and the Netherlands that would be evident should other forms of mobility, such as short-term stays, also be considered.

In terms of explanations for the result, if taken at face value, factors associated with the different institutional conditions characterizing the two national systems need to be taken into account. First, the Spanish and Dutch university and research systems are different in many ways (Deloitte 2014a, 2014b; IDEA Consult 2012a, 2012b). For example, the Netherlands higher education system appears to be more internationalized than that of Spain. A higher proportion of PhD students enrolled in universities in the Netherlands are international students, compared to the situation in Spanish universities (OECD 2015). Data for 2013 show that 38% of doctoral students in the Netherlands were international students, compared to 16% in Spain and an OECD average of 24%. At the Masters level, 17% of students in the Netherlands were international students, compared to 5% in Spain and an OECD average of 14% (OECD 2015: 354). Overall, the Netherlands was hosting 2% of the global population of international students and Spain 1% (OECD 2015).

If we assume that many early career researchers will publish their first paper either during their PhD studies or shortly thereafter, then this first paper is very likely to be affiliated with the training university. In the Dutch case then, a higher proportion of these papers will be published by international students/researchers. If we then also assume that a substantial proportion of these international students will obtain a position in their home country, then the proportion of the 'outward' mobility from the Netherlands that is also in effect return migration by international students, will likely be higher than in the case of Spain. In other words, the more internationalized nature of the Dutch higher education system means that a higher proportion of the initial mobility included in this study is actually already part of a chain of moves. Disentangling the extent to which initial career mobility outward mobility of nationals from the training country or returning internationals is a topic for future research. At this stage we should be very cautious to not interpret our initial results as reflecting particularly Dutch or Spanish propensities for being mobile researchers. However, what seems likely is that the degree of internationalization that characterizes the research systems of the Netherlands and Spain may well include different levels of specific forms of scientific mobility, which have a different role in the research careers of the researchers concerned.

Second, there may be very different institutional incentives for mobility in the two systems. Cruz-Castro and Sanz-Menendez (2010) studied the interrelations between research mobility, productivity performance and tenure in Spain. They found that there was no return to careers from mobility in terms of accelerated progress to tenure. Due to labour market and other institutional factors associated with employment in Spain, researchers who stay within their PhD department may be advantaged through increased productivity and lower transaction costs associated with trying to access a permanent position. It is evident that mobility is not necessarily beneficial to academic careers in all respects. In fact, mobility may even be associated with a delay in tenure in the Spanish case (Sanz-Menendez et al. 2013). Whilst mobility may still have considerable value for Spain-based researchers, this may be primarily associated with internal evaluations and promotion processes, thereby creating incentives for types of mobility that do not interrupt job access and continuity within a Spanish institution. It is not directly clear how this compares to the case of the Netherlands – although the different institutional conditions that exist there (Deloitte 2014a; IDEA Consult 2012a) suggests the outcome of mobility, performance and tenure processes will produce different results to those observed in Spain. This is an area for further national and comparative research.

Another area of potential future research concerns the prevalence of researchers with multiple affiliations in the sample groups in both the Netherlands and Spain. One in ten researchers based in the Netherlands has a multiple affiliation. This *may* be linked to the internationalization of the Dutch system. However, at this stage there is no clear evidence to support this assumption. In the Spanish case the rate of multiple affiliations is only slightly lower. This may be linked to the recent wave of 'forced emigration' of researchers from Spain due to the severity of the financial crisis and its impact on research funding, coupled with these researchers' desire to maintain links with 'home', and potential 'return', organisations. However, evidence is lacking and these comments are largely speculative. More sophisticated bibliometric analyses and comparisons with different 'generation' of researchers, possibly combined with qualitative investigation, will be required to better understand the role and function of multiple affiliations in both research mobility and research careers.